\documentclass[sigconf]{acmart}

\AtBeginDocument{%
  \providecommand\BibTeX{{%
    \normalfont B\kern-0.5em{\scshape i\kern-0.25em b}\kern-0.8em\TeX}}}

\copyrightyear{2024}
\acmYear{2024}
\setcopyright{rightsretained}
\acmConference[IDE '24]{2024 First IDE Workshop}{April 20, 2024}{Lisbon, Portugal}
\acmBooktitle{2024 First IDE Workshop (IDE '24), April 20, 2024, Lisbon, Portugal}\acmDOI{10.1145/3643796.3648459}
\acmISBN{979-8-4007-0580-9/24/04}

\usepackage{subfigure}
\usepackage{balance}
\begin{document}

\title{Gamified GUI testing with Selenium in the IntelliJ IDE: A Prototype Plugin}

\author{Giacomo Garaccione}
\email{giacomo.garaccione@polito.it}
\affiliation{%
  \institution{Politecnico di Torino}
  \city{Turin}
  \country{Italy}
}

\author{Tommaso Fulcini}
\email{tommaso.fulcini@polito.it}
\affiliation{%
  \institution{Politecnico di Torino}
  \city{Turin}
  \country{Italy}
}

\author{Paolo Stefanut Bodnarescul}
\affiliation{%
  \institution{Politecnico di Torino}
  \city{Turin}
  \country{Italy}
}

\author{Riccardo Coppola}
\email{riccardo.coppola@polito.it}
\affiliation{%
  \institution{Politecnico di Torino}
  \city{Turin}
  \country{Italy}
}

\author{Luca Ardito}
\email{luca.ardito@polito.it}
\affiliation{%
  \institution{Politecnico di Torino}
  \city{Turin}
  \country{Italy}
}

\renewcommand{\shortauthors}{Garaccione et al.}

\begin{abstract} 
Software testing is a crucial phase in software development, enabling the detection of issues and defects that may arise during the development process. Addressing these issues enhances software applications' quality, reliability, user experience, and performance. Graphical User Interface (GUI) testing, one such technique, involves mimicking a regular user's interactions with an application to identify defects. However, GUI testing is often underutilized due to its perceived repetitiveness, error-proneness, and lack of immediate feedback on test quality.
In recent years, gamification—incorporating game elements in non-game contexts to boost interest, motivation, and engagement—has gained traction in various fields, including software engineering and education. This paper presents GIPGUT: a prototype of a gamification plugin for IntelliJ IDEA, an Integrated Development Environment (IDE) that supports scripted GUI testing. The plugin enhances testers' engagement with typically monotonous and tedious tasks through achievements, rewards, and profile customization.
A preliminary prototype evaluation was conducted with a small group of users to assess its usability and the impact of gamification on the GUI testing process. The results indicate high usability and positive reception of the gamification elements. However, due to the limited sample size of participants, further research is necessary to understand the plugin's effectiveness fully.

\end{abstract}

\ccsdesc[500]{Software and its engineering~Software testing and debugging}
\ccsdesc[500]{Software and its engineering~Integrated and visual development environments}

\keywords{software testing, graphical user interface testing, gamification, integrated development environment, web testing}

\received{7 December 2023}

\maketitle

\section{Introduction}
Software testing is a critical phase in software development, aiming to identify issues and problems in a software product before its final release. Fixing these issues during testing enhances the quality and performance of the final product.

Among various testing methodologies, Graphical User Interface (GUI) testing examines the behaviour of an entire application by simulating a user's interactions to identify issues and defects. Despite its importance, GUI testing is often neglected in practice, being perceived as a repetitive, error-prone activity that needs immediate feedback on test quality.

Gamification, defined as ``the use of game-like elements in non-game contexts to increase interest, motivation, and participation" \cite{inproceedings:gamification}, has been widely adopted in software engineering with promising results. This includes its recent application in GUI testing.

This paper explores the application of gamification to GUI testing by describing a prototype for IntelliJ IDEA, a well-known Integrated Development Environment (IDE) that supports scripted GUI testing. The plugin introduces mechanics aimed at enhancing testers' interest in testing, such as achievements, profile customization, daily tasks, level progression, and unlockable content.

The remainder of this paper is structured as follows: Section \ref{sec:background} presents relevant background information on GUI testing and gamification. Section \ref{sec:design} describes the plugin, detailing its implementation and the gamification mechanics it incorporates. Section \ref{sec:evaluation} describes a preliminary evaluation of the plugin. Finally, Section \ref{sec:end} discusses the current limitations of our work and outlines plans for future development of the plugin.

\section{Background}
\label{sec:background}
Graphical User Interface (GUI) testing involves assessing the behaviour of a complete application by interacting with its UI, as a regular user would. This includes checking for the presence and positioning of specific elements, ensuring the functionality and accessibility of different elements, and verifying the correct handling of error scenarios.

The most prevalent method for GUI testing is manual, where a tester executes a sequence of operations as per a test case. However, recent trends show a shift towards automated testing. In this approach, test cases are not executed manually but are defined as automated scripts, thus reducing time consumption and tedium. Automated GUI testing strategies include scripted GUI testing (using dedicated frameworks and APIs for script writing), Capture and Replay (generating scripts from recorded GUI interactions), and Automated Test Generation (employing random or systematic inputs to navigate the user interface).

Automated GUI testing is beneficial as it requires fewer developers and less time than manual testing. It also enhances bug detection and test coverage \cite{article:literature_web_testing}. Nevertheless, a significant drawback is the fragility of test scripts: GUI changes can cause previously functional scripts to fail, necessitating developer intervention for script updates.

An example of a scripted automated GUI testing tool is Selenium WebDriver. This open-source tool automates web browsers and supports testing operations like acceptance, functional, performance, load, and stress testing.

Selenium WebDriver simulates a web browser and offers APIs in multiple programming languages for interacting with a web application’s GUI. This flexibility allows developers to choose the most suitable language for their needs. Testers can identify widgets on a web page using `locators' — unique properties such as id, name, CSS class name, tag name, or XPath that pinpoint specific page elements. However, test cases created with WebDriver may fail if the locators are not updated following changes in the system under test.

Gamification has emerged as a popular strategy in recent years to boost interest and motivation in various contexts, both educational \cite{dubois, uskov} and industrial \cite{PEDREIRA2015157, gamif_diversity}.

Numerous frameworks have been established in the field of gamification. Among them, one of the most prominent, and upon which this article is based, is Octalysis \cite{octalysis}. The key appeal of Octalysis lies in its `human-focused design,' as termed by the creator. This approach prioritizes understanding and catering to the user's emotions, motivations, and interests rather than emphasizing efficiency or rapid results.

\begin{figure}
    \centering
    \includegraphics[width=0.7\columnwidth]{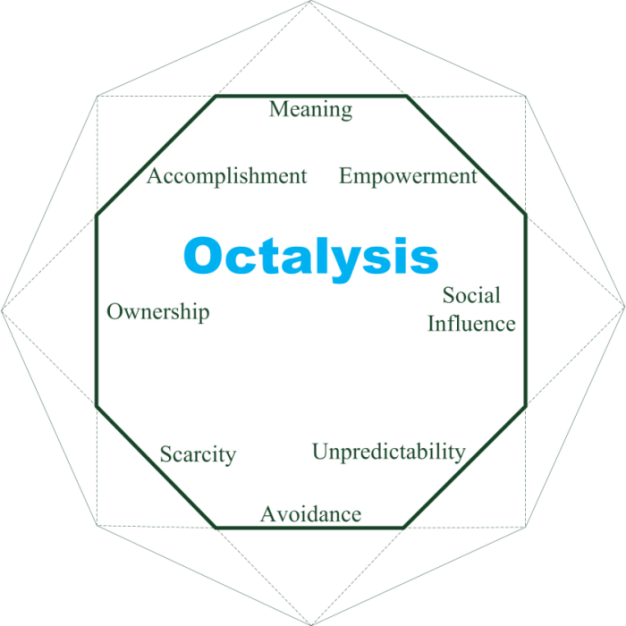}
    \caption{Octalysis diagram. (Source: Li \cite{li2019gamification})}
    \label{fig:octalysis}
\end{figure}

The Octalysis framework delineates eight specific `Core Drives' that encapsulate various human aspects activated by gamification. These drives are the underlying reasons that motivate users to engage in certain activities and contribute to the enjoyment of games. Each drive represents a distinct aspect of human behaviour and is illustrated in Figure \ref{fig:octalysis}.

Octalysis incorporates two primary distinctions within its core drives. The first is the division between `White Hat' and `Black Hat' drives. White Hat drives utilize positive motivators, such as giving players a sense of control over their actions and satisfaction with the outcomes. Conversely, Black Hat drives leverage more negative motivators, creating feelings of anxiety and exploiting fears such as the unknown or the fear of loss.

The second distinction differentiates between `intrinsic' and `extrinsic' motivators. Intrinsic motivators are associated with the right side of the brain, relating to creativity, self-expression, and socialization. These motivators are fulfilling in themselves, without any specific external reward. On the other hand, extrinsic motivators are linked to the left side of the brain and are driven by the pursuit of specific goals or rewards.

An effective gamified system, adhering to the Octalysis framework, should offer a balanced experience encompassing both sets of distinctions. A system focused solely on extrinsic, tangible rewards and positive motivators is not considered complete, as it needs certain elements to enhance the overall experience.

Currently, the literature about gamification in software testing is mainly related to unit testing, the use in GUI testing is secondary \cite{MLR}. One notable framework, developed by Cacciotto et al. \cite{cacciotto}, incorporates elements such as progress indicators to show page coverage, easter eggs for discovering new web pages, points awarded for identifying new page elements, and leaderboards to foster a competitive environment. Such a framework has been empirically validated, proving that gamification can effectively drive testers' behaviour during their work \cite{10375892}.

Another example is GERRY \cite{gerry}, a Google Chrome extension that adheres to the principles outlined in the framework mentioned above and introduces additional features. These include enabling testers to report issues and the automated generation of test scripts compatible with prevalent GUI testing tools.

\section{Plug-in Design}
\label{sec:design}
In this section, we detail our plugin's design and implementation aspects. Specifically, Section \ref{sec:implementation} outlines the design and implementation process, while Section \ref{sec:gamification} discusses the gamification mechanics aligned with the core drives of the Octalysis framework.

\subsection{Implementation}
\label{sec:implementation}

The original idea of a gamification plugin based on achievements was presented by Straubinger et al. \cite{straubinger2023improving}. They created an achievements-based gamification plugin for IntelliJ IDEA to assist testers in executing their test using the JUnit framework. The authors report as a result of their experiment that the involved testers wrote better tests and tested their code more frequently when their plugin was activated. Based on this work, we decided to develop a similar plugin, which instead of providing achievements based on the performance of unit testing, focuses on web testing at the GUI level.

During the design phase, we evaluated several Integrated Development Environments (IDEs), including Visual Studio Code, Eclipse, JetBrains' IntelliJ IDEA, and JetBrains Aqua, ultimately selecting IntelliJ IDEA for our plugin development.

IntelliJ IDEA was chosen for its significant features, such as cross-language support, intelligent code assistance, and a robust plugin ecosystem. In contrast, the alternative options presented various challenges: Visual Studio Code exhibited compatibility issues with Java, Eclipse had a steeper learning curve, and JetBrains Aqua was still in preview mode with limited support.

\begin{figure}[ht]
\centering
\includegraphics[width=\columnwidth]{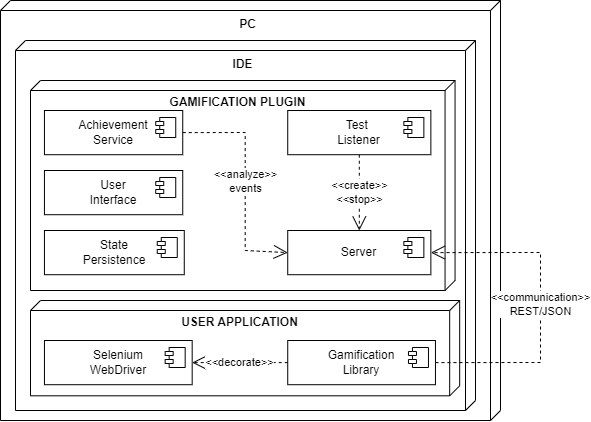}
\caption{UML Deployment Diagram representing the plugin infrastructure}
\label{fig:deploy}
\end{figure}

The plugin comprises two distinct modules: the gamification plugin, developed as an IntelliJ IDEA extension, and the user application where the Selenium WebDriver conducts its tests. Figure \ref{fig:deploy} shows a deployment diagram representing the two modules.

This separation into two distinct modules was necessary due to a challenge encountered during development: obtaining data from the Selenium WebDriver during testing was not feasible using only the IntelliJ Platform. The APIs provided by the IntelliJ Platform are designed for static analysis rather than for use during program execution. Consequently, we developed an external library with a listener embedded in the WebDriver. This listener executes tasks before and after each WebDriver call, enabling data collection to be sent to the plugin at the end of a test. This approach was chosen as the most effective compromise to minimize the setup complexity for end-users, which at present time is limited to the inclusion of the library to the project and the binding between the used WebDriver and the library.

The two modules communicate via a local server initiated by the plugin before testing. This server remains active throughout the testing process, listening to each test to facilitate the generation of necessary statistics. Upon completing a WebDriver test, the events associated with the test are transmitted to the `Achievement Service' via the server. The Achievement Service then calculates the progress towards each achievement, a concept elaborated upon in the following subsection.

Persistence across different testing sessions is vital for tracking user progression. This is achieved with the `State Persistence' class, which stores essential user profile information, including levels and achievements.

The external library was developed based on Selenium’s 'WebDriverInterface', which provides listeners for web events such as page loads or element clicks. However, a significant issue arose during development: choosing the appropriate locator to identify elements interacted with by the WebDriver’s various methods. Options included unique ID, name, XPath, or CSS style. Our strategy was to select the first non-null option among ID, name, and XPath. This decision was made because not all developers adhere to uniform conventions, resulting in some web page elements lacking an ID or name. However, despite being considered rather a fragile identifier, an XPath is always available. \cite{test-smell}

\subsection{Gamification Mechanics}
\label{sec:gamification}
This section outlines the various gamification mechanics implemented in our plugin, aligned with the eight Core Drives identified in the Octalysis framework. The selected mechanics are as follows:

\begin{enumerate}
\item \textbf{Profile Customization.} The plugin offers users a dedicated profile page that can be personalized by altering the displayed username, icon, title, and visible achievements. This customization enhances the sense of ownership and identity, providing each user with a unique experience and encouraging ongoing interaction with the plugin. The availability of rare content for profiles, which users can unlock and showcase, fosters a deeper connection between the user and their achievements. Profile customization is linked to the `Ownership' Core Drive in the Octalysis framework and is a widely used mechanic in various gamification frameworks, as noted by Pedreira et al. \cite{gamif_architecture}. A visual representation of the profile section is provided in Figure \ref{fig:profile}.

\begin{figure}
    \centering
    \begin{minipage}{0.48\columnwidth}
        \centering
        \includegraphics[width=\textwidth]{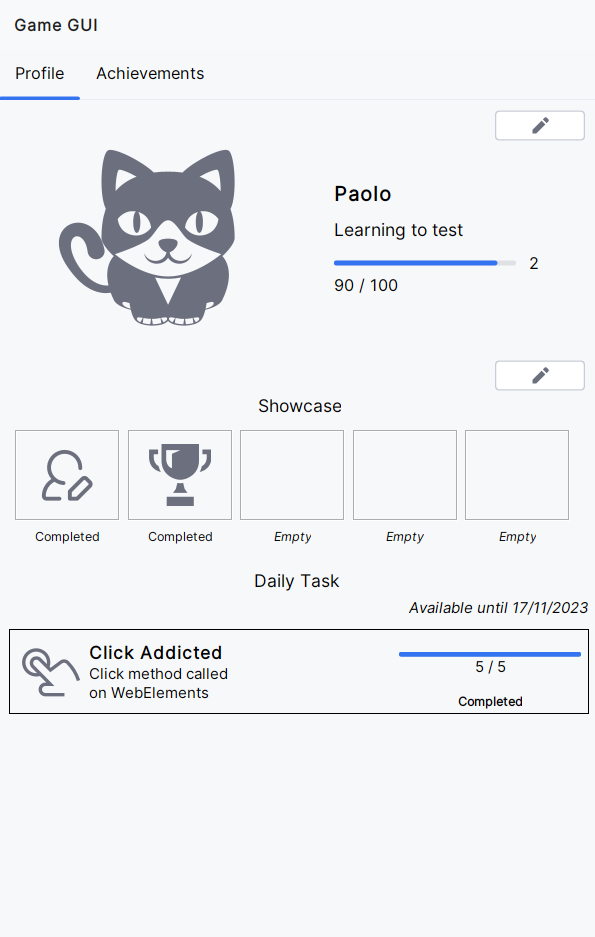} 
        \caption{Profile section}
        \label{fig:profile}
    \end{minipage}\hfill
    \begin{minipage}{0.48\columnwidth}
        \centering
        \includegraphics[width=\textwidth]{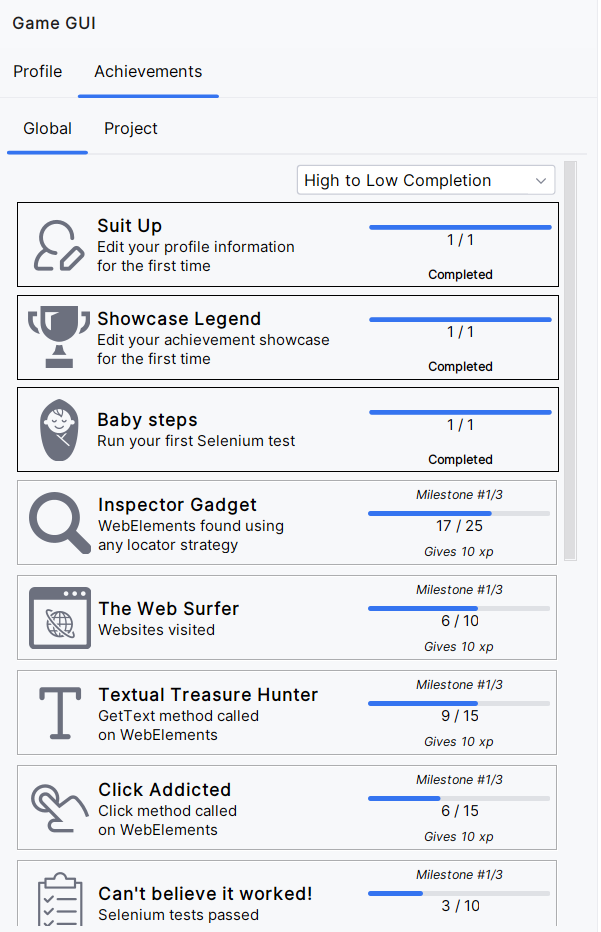} 
        \caption{Achievements}
        \label{fig:achievements}
    \end{minipage}
\end{figure}

\item \textbf{Progression.} The profile page features a level indicator and a progress bar, representing the user's advancement towards the next level. This level is a numerical symbol of their skill as a tester and measures their accomplishments and interactions with the plugin. The constant visibility of the level also acts as a motivator, encouraging users to take on new challenges and grow in their testing abilities. Users begin at level 1 and accumulate experience towards higher levels by completing achievements within the tool, with the highest attainable level being 10. This progression mechanism embodies the `Accomplishment' Core Drive in gamification. Level progression is a ubiquitous element in gamification theory, recognized for its simplicity in implementation and effectiveness \cite{garcia2017framework, gamif_architecture, bell2011secret}.

\item \textbf{Unlockable Content.} The plugin's unlockable content includes user icons and titles. Icons are images that users can display on their profiles, while titles are short phrases accompanying the user's name, reflecting their status. Both icons and titles are linked to the user's level progression; new options unlock at each level, providing a tangible sense of progression and achievement. Rarer icons and titles at higher levels carry greater intrinsic value, motivating users to engage more thoroughly in testing to earn these coveted rewards. The `Ownership' Core Drive is associated with this unlockable content. Such rewards are effective motivators in numerous gamified tools \cite{costa2019systematic, gamif_architecture}

\item \textbf{Achievements.} Achievements are a pivotal gamification element in the plugin, as levels, experience, and unlockable content are all linked to obtaining achievements. An icon, a name, and a description denote each achievement. The icon and name are designed to be memorable and easily recognizable, while the description clarifies the activities required to earn the achievement. Achievements have various milestones, rewarding experience points upon completion. 

They are categorized into global achievements, which pertain to progress across the entire application and have higher milestones and project achievements specific to an IntelliJ project. Some global achievements mirror project achievements but are more challenging. Figure \ref{fig:achievements} illustrates the achievements section in the plugin. Most achievements award the user based on GUI-related behaviours, i.e. actions performed with Selenium WebDriver, such as website visits or WebElement interactions. Others recognize more general accomplishments, like fixing failed tests or rewarding users for initial profile customization.

Additionally, users can showcase up to five achievements on their profile, highlighting their most significant or recent successes. However, currently, there is no feature for users to view each other’s profiles, limiting profile sharing to external means, such as screenshots. Achievements align with the `Accomplishment' Core Drive, while the achievement showcase relates to the `Social Influence' Drive. Like levels, achievements are a common aspect in gamification \cite{bell2011secret, sheth2013competitive}.

\item \textbf{Daily Tasks.} A daily task is a time-limited achievement, selected randomly each day to maintain user engagement and offer fresh challenges, even to those who have completed all available achievements. Daily tasks resemble global achievements but have lower completion thresholds. They embody the `Unpredictability' Core Drive. While not as prevalent as other mechanics, daily tasks are an innovative feature; Coppola et al. \cite{Coppola2023} provide an example of incorporating daily tasks into a gamification framework.

\end{enumerate}

\section{Preliminary Evaluation}
\label{sec:evaluation}
A preliminary evaluation of the gamification plugin was conducted following its development. This assessment aimed to gauge the reception of the gamified mechanics, identify potential issues, and pinpoint areas for improvement.

The participants in the experiment were four students, each holding a Master's degree in Software Engineering and possessing experience in Object-Oriented Programming and web development.

The evaluation consisted of three tasks to be completed within a maximum time frame of 45 minutes. Before the experiment, participants received a brief introduction to GUI testing and the Selenium WebDriver. This introduction was designed to equip them with the necessary knowledge to complete the tasks. Following the session, participants were asked to complete a survey to provide feedback and evaluate the plugin's features and usability.

The tasks were centred around the popular e-commerce website Amazon, serving to analyze user interaction with the plugin and the impact of gamification on their approach to GUI testing. The tasks, designed to have specific objectives but allowing freedom in execution, were as follows:

\begin{enumerate}
\item The tester was required to search for a specific product on the website, select the first result and verify that its price was below a predetermined threshold.
\item Building on Task 1, the tester had to navigate to the product’s review page, ensuring the accuracy of the title and author for both the top positive and critical reviews.
\item Continuing from Task 1, the tester needed to add the product to the cart, proceed to the cart page, and confirm the product's name, quantity, price, and subtotal. Then, the tester had to visit the checkout page and check for the presence of a button to create a new account.
\end{enumerate}

Among the four participants, only one was unable to complete all three tasks within the allotted time, struggling to locate the required page elements. However, all participants achieved enough to reach level 2, enabling them to experience the full range of mechanics offered by the plugin, including achievements, titles, and showcasing their attained achievements.

The plugin's usability was assessed using the System Usability Scale (SUS) \cite{article:sus}, a widely recognized questionnaire for evaluating a system's perceived usability. It's important to note that the survey focused solely on the usability of the plugin rather than on the combined usability of IntelliJ IDEA and the plugin.

The SUS questionnaire results were predominantly positive, with most questions receiving favourable responses. The final SUS score was \textbf{93.75}, significantly surpassing the threshold of 68 required for a system to be considered usable. However, it should be acknowledged that SUS scores are not absolute usability indicators. Despite this caveat, the high usability score of the plugin is a promising outcome, particularly as usability is a critical factor for the success of a gamified system.

\section{Limitations and Future Work}
\label{sec:end}
In this paper, we presented a gamification plugin for IntelliJ IDEA designed to enhance test script generation for GUI testing using the Selenium WebDriver. The plugin, evaluated in an experiment with four participants, demonstrated promising results in usability and user appreciation.

However, the plugin is still in its prototype stage and faces several limitations. Firstly, the necessity of including a gamification library for tracking actions via the Selenium WebDriver is a constraint, although the overhead is limited to the inclusion of the library jar file to the build path of the project. 

Additionally, the scope of trackable actions is limited to those compatible with the WebDriverListener interface. Another challenge is the reliance on external XML files for data persistence. These files are susceptible to modification or deletion by the user and can potentially slow down the plugin if overloaded with data. The use of simple XML files also makes the use of the plugin a purely single-player experience at the moment, without allowing users to view colleagues' profiles. Future works should focus on making the gaming experience shared between participants, relying on external sources for data persistence.

Furthermore, the small participant sample in our initial evaluation limits our ability to conclusively assess the gamification mechanics' impact. A larger-scale experiment is required to determine whether the plugin genuinely enhances tester efficiency and morale. Such a study would also provide insights into the efficacy of the gamification mechanics for GUI testing and whether they require further refinement.

The source code is publicly available in the project repository:\\
{
\centering \url{https://github.com/SoftengPoliTo/GIPGUT}

}
\balance

\section*{Acknowledgement}

This study was carried out within the “EndGame - Improving End-to-End Testing of Web and Mobile Apps through Gamification” project (2022PCCMLF) – funded by European Union – Next Generation EU within the PRIN 2022 program (D.D.104 - 02/02/2022 Ministero dell’Università e della Ricerca). This manuscript reflects only the authors’ views and opinions and the Ministry cannot be considered responsible for them.

\bibliographystyle{ACM-Reference-Format}
\bibliography{selenium}
\end{document}